\documentclass[%
 reprint,
 amsmath,amssymb,
 aps,
]{revtex4-2}

\usepackage{graphicx}
\usepackage{dcolumn}
\usepackage{bm}
\usepackage{physics}
\usepackage{hyperref}
\usepackage[mathlines]{lineno}

\usepackage{color}

\begin{document}

\preprint{APS/123-QED}

\title{
Magnetism and superconductivity in the $t-J$ model of $La_3Ni_2O_7$ \\
under multiband Gutzwiller approximation}

\author{Jie-Ran Xue}
\affiliation{%
International Center for Quantum Materials, School of Physics, Peking University, Beijing 100871,
China
}%

\author{Fa Wang$^{\ast}$}
\email{wangfa@pku.edu.cn}
\affiliation{%
International Center for Quantum Materials, School of Physics, Peking University, Beijing 100871, China and \\
Collaborative Innovation Center of Quantum Matter, Beijing 100871, China
}%
\date{\today}

\begin{abstract}
The recent discovery of 
possible high temperature
superconductivity in single crystals of $La_3Ni_2O_7$ under pressure renews the interest in research on nickelates.
The DFT calculations reveal that both $d_{z^2}$ and $d_{x^2-y^2}$ orbitals are active, which suggests a minimal two-orbital model to capture the low-energy physics of this system.
In this work, we study a bilayer 
two-orbital 
$t-J$ model within multiband Gutzwiller approximation, and discuss the 
magnetism
as well as the superconductivity over a wide range of the hole doping.
Owing to the inter-orbital super-exchange process between $d_{z^2}$ and $d_{x^2-y^2}$ orbitals, the induced ferromagnetic coupling within layers competes with the conventional antiferromagnetic coupling, and leads to complicated hole doping dependence for the magnetic properties in the system.
With increasing hole doping, the system transfers to A-AFM state from the starting G-AFM state.
We also find the inter-layer superconducting pairing of $d_{x^2-y^2}$
orbitals 
dominates due to the large hopping parameter of $d_{z^2}$ along the vertical inter-layer bonds and significant Hund's coupling between $d_{z^2}$ and $d_{x^2-y^2}$ orbitals.
Meanwhile, the G-AFM state and superconductivity state can coexist in the low hole doping regime.
To take account of the pressure, we also analyze the impacts of inter-layer hopping amplitude on the system properties.
\end{abstract}

\maketitle

\section{\label{sec:level1}Introduction}
Since the discovery of 
high-$T_c$ superconductivity in cuprates, 
extensive experimental and theoretical efforts have been made to explain the pairing mechanism in these compounds.
Most of theoretical studies suggested 
that the $d^9$ electronic configuration of $\text{Cu}^{2+}$ and quasi-two-dimensional $\text{Cu}\text{O}_2$ layers play critical roles in cuprate superconductivity.
Considering $\text{Ni}^{+}$ is isoelectronic with $\text{Cu}^{2+}$, the search for unconventinal superconductivity in nickelates 
has become 
one major focus in condensed matter community \cite{PhysRevB.59.7901,Li2019,Osada2020,zeng2022superconductivity,PhysRevLett.125.027001,PhysRevLett.125.027001,
PhysRevMaterials.4.121801,Zeng2022,doi:10.1021/acs.nanolett.3c00192,Wang2022,GU2022100202,Pan2022}.
Although superconductivity has been discovered in several nickelates, such as hole-doped infinite-layer nickelates $\mathrm{Nd}_{1-x}\mathrm{Sr}_x\mathrm{Ni}\mathrm{O}_2$, 
the maximum of transition temperature $T_c$ is just over $30K$ and superconductivity only appears in thin films.
Recently, 
evidences of 
superconductivity were observed in single crystal $\mathrm{La}_3\mathrm{Ni}_2\mathrm{O}_7$ under pressure, and the transition temperature $T_c$ can reach $80K$\cite{Sun2023}.
This experimental discovery immediately attracts great attention, and provides a new platform for researches of high-$T_c$ superconductivity.

According to the Zhang-Rice singlets picture\cite{PhysRevB.37.3759}, the low-energy physics of cuprates can be well described by a one band Hubbard model, while situation becomes more complex in nickelates.
The average valence of $\mathrm{Ni}$ in $\mathrm{La}_3\mathrm{Ni}_2\mathrm{O}_7$ is $d^{8-\delta}$ with $\delta=0.5$
for the superconducting samples, 
and density functional theory (DFT) calculations reveal that both the $d_{z^2}$ orbital and $d_{x^2-y^2}$ orbital contribute near the Fermi level
\cite{Sun2023,PhysRevLett.131.126001,gu2023effective,PhysRevB.83.245128}.
Owing to the apical oxygen between Ni-O bilayers, the $d_{z^2}$ orbitals has strong inter-layer hopping and can form bonding-antibonding molecular-orbital states, results in energy splitting between $d_{z^2}$ and $d_{x^2-y^2}$ bands. 
The above composite scenario suggests a two-orbital minimal model, where $d_{z^2}$ is half-filled and $d_{x^2-y^2}$ is quarter-filled. 
We believe that two electrons at the same site prefer to form high-spin ($S=1$) configuration, since the Hund's coupling, denoted as $J_H$ hereafter, is much larger than the splitting energy between $d_{z^2}$ and $d_{x^2-y^2}$ orbitals\cite{PhysRevLett.131.206501,cao2023flat,kumar2023softening}.
In the large $J_H$ limit, we also include the intra-orbital and inter-orbital Hubbard interactions, and derive an effective bilayer $t-J$ model to describe the low-energy physics of $\text{La}_3\text{Ni}_2\text{O}_7$.
Similar model has been used to study the superconducting nickelate $Nd_{1-x}Sr_xNiO_2$ \cite{PhysRevResearch.2.023112}.

There are already many theoretical
\cite{PhysRevLett.131.126001,PhysRevLett.131.206501,wu2023charge,
PhysRevLett.132.036502,PhysRevB.108.L201108,lu2023superconductivity,schlomer2023superconductivity,chen2023orbitalselective,
qu2023roles,lu2023interlayer,zheng2023superconductivity,
lu2023interplay,PhysRevLett.131.236002,PhysRevB.108.L140505,PhysRevB.108.165141,sakakibara2023possible,gu2023effective,zhang2023structural,ryee2023critical,zhang2023electronic,
PhysRevB.109.045127,yang2023strong,
PhysRevB.108.174511,kakoi2023pair,yang2023effective,
jiang2023pressure,heier2023competing,fan2023superconductivity,
PhysRevB.108.L180510,labollita2023electronic,chen2023critical,
kumar2023softening,sui2023electronic,geisler2024optical} 
and experimental studies
\cite{zhang2023hightemperature,zhou2023evidence,wang2023,wang2023longrange,liu2023electronic,kakoi2023multiband,chen2023evidence,
xu2023pressuredependent,dong2023visualization,talantsev2024debye}
for the bilayer nickelate $La_3Ni_2O_7$. 
Most of these works focus on the pairing mechanism and pairing symmetry in the superconducting phase.
Some studies emphasize the similarity of the electronic structures between nickelates and cuprates, 
and
further suggest that 
the strong hybridization between O $2p$ orbitals and Ni $3d$ orbitals
will lead
to the emergence of Zhang-Rice singlets and 
$d$-wave pairing in nickelates\cite{jiang2023pressure}.
Other works believe that the $d_{z^2}$ orbitals plays an important role in these nickelates.
Within this class of works, some researchers believe that the enhancement of 
the inter-layer hopping of $d_{z^2}$ orbitals under
pressure can induce the metallization of the energy band originating from $d_{z^2}$ orbitals\cite{Sun2023,yang2023effective}, which is beneficial to superconductivity.
Some other researchers have further studied 
the interaction between $d_{z^2}$ and $d_{x^2-y^2}$ orbitals, and 
derived various multiband models for these materials.
Some of these models are predicted to produce $s_{\pm}$ pairing symmetry
\cite{PhysRevLett.131.236002,PhysRevB.108.L140505,PhysRevB.108.165141,sakakibara2023possible,gu2023effective,zhang2023structural,ryee2023critical,zhang2023electronic}.
There are also a few theoretical works about possible magnetism in 
$\mathrm{La}_3\mathrm{Ni}_2\mathrm{O}_7$ 
\cite{chen2023critical,PhysRevB.108.L180510,labollita2023electronic}
However the interplay between magnetism and superconductivity, which is an extremely important issue in Cu- and Fe-based high-$T_c$ superconductivity, has not been carefully studied in the context of $\mathrm{La}_3\mathrm{Ni}_2\mathrm{O}_7$.

In this work, we study a bilayer $t-J$ model 
for $\mathrm{La}_3\mathrm{Ni}_2\mathrm{O}_7$ 
using multiband Gutzwiller approximation, 
to have a comprehensive understanding of the ground state properties of this system at different band filling.
We find that inter-layer 
$s$-wave 
superconducting pairing of $d_{x^2-y^2}$ orbitals dominates and coexists with 
G-type antiferromagnetic(G-AFM) 
order in the low doping region. 
Meanwhile, superconductivity can be enhanced with increasing inter-layer hopping amplitude.
With increasing hole doping, the system tranfers to 
A-type antiferromagnetic(A-AFM) 
state. 
The rest of the paper is as follows. In Sec.~\ref{sec:level2}, we introduce the minimal two-orbital model and the derived bilayer $t-J$ Hamiltonian. 
Then, we describe the procedures of performing the multiband Gutzwiller approximations and determining the renormalized mean-field Hamiltonian in detail.
In Sec.~\ref{sec:level3}, we present the numerical results obtained from solving the renormalized mean-field Hamiltonian in a self-consistent manner, and discuss the interplay of superconductivity, antiferromagnetism, as well as ferromagnetism in a wide range of hole doping. 
Finally, we present our conclusions 
and comparison with previous studies
in Sec.~\ref{sec:level4}.

\section{\label{sec:level2}Bilayer $t-J$ model and\protect\\ Effective Mean-Field 
Hamiltonian}
\begin{figure}[h]
\includegraphics[width=\columnwidth]{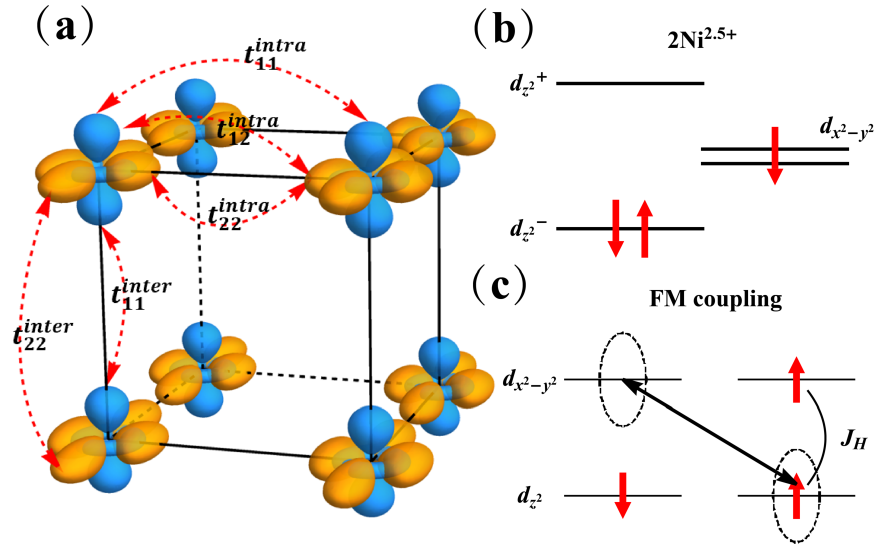}
\caption{\label{fig:Fig1} (a) The schematic of the bilayer two-orbital model with hopping parameters in the Hamiltonian. 
(b) Electronic configuration in two atoms with the same planar site, where $d_{z^2}^+$ represents the anti-bonding state and $d_{z^2}^-$ is the bonding states.
(c) The inter-orbital superexchange process which induces the ferromagnetic coupling.
}
\end{figure}

We start from a two-orbital (Ni$-d_{z^2}$ and Ni$-d_{x^2-y^2}$) model on a bilayer square lattice(as depicted in Fig.~\ref{fig:Fig1}):\\
\begin{eqnarray}
H&&=H_{t}+\frac{U_1}{2}\sum_{il}n_{1l;i}(n_{1l;i}-1)+\frac{U_2}{2}\sum_{il}n_{2l;i}(n_{2l;i}-1)\nonumber\\
&&+U^{\prime}\sum_{il}n_{1l;i}n_{2l;i}-2J_{H}\sum_{il}(\vec{S}_{1l;i}\cdot\vec{S}_{2l,i}+\frac{1}{4}n_{1l;i}n_{2l;i}),\nonumber\\
H_{t}&&=\sum_{il\alpha}\epsilon_{\alpha}n_{\alpha l;i}+\sum_{\langle ij\rangle}\sum_{l\sigma \alpha\beta}t^{\text{intra}}_{\alpha\beta}c^{\dagger}_{\alpha l;i\sigma}c_{\beta l;j\sigma}\nonumber\\
&&+\sum_{il\sigma\alpha\beta}t^{\text{inter}}_{\alpha\beta}c^{\dagger}_{\alpha l;i\sigma}c_{\beta \bar{l};i\sigma},
\end{eqnarray}\noindent
where $l=t,b$ labels the top and bottom layers, $\alpha/\beta=1,2$ denotes the two orbitals $d_{z^2}$ and $d_{x^2-y^2}$ respectively, and $\sigma=\uparrow, \downarrow$ labels the spin.
$\langle ij\rangle$ denotes the nearest-neighbor within each layer. 
$n_{\alpha l;i}=\sum_{\sigma}c^{\dagger}_{\alpha l;i\sigma}c_{\alpha l;i\sigma}$ is the particle number operator of orbital $\alpha$ at site $i$ within layer $l$,
and $\vec{S}_{\alpha l;i}=\frac{1}{2}\sum_{\sigma \sigma^{\prime}}c^{\dagger}_{\alpha l;i\sigma}\vec{\sigma}_{\sigma\sigma^{\prime}}c_{\alpha l;i\sigma^{\prime}}$ is the spin operator.
We consider the onsite intra-orbital Hubburd interaction $U_1$, $U_2$, inter-orbital repulsion $U^{\prime}$ and Hund's coupling $J_H$.
We assume $U_1=U_2=U$ and adopt Kanamori relation $U=U^{\prime}+2J_H$\cite{PhysRevB.18.4945}. $\epsilon_{\alpha}$ is the onsite energy and reflects the splitting energy between the two orbitals.
According to DFT calculation\cite{PhysRevLett.131.126001}, we take $t^{\text{intra}}_{11}=-0.11$, $t^{\text{intra}}_{22}=-0.483$, $t^{\text{intra}}_{12}=0.239$, $t^{\text{inter}}_{22}=0.005$, $t^{\text{inter}}_{12}=0$, $\epsilon_0=0$ and $\epsilon_1=0.367$. 
Meanwhile we set $U=4$, $U^{\prime}=2$, and $J_H=1$\cite{PhysRevLett.131.206501,chen2023critical,cao2023flat,kumar2023softening}. Varying pressure, the inter-layer hopping amplitude for orbital $d_{z^2}$ may significantly increase\cite{zhang2023structural}, and our work considers different values of $t^{\text{inter}}_{11} (=-0.5, -0.6, -0.7)$. In addition, the average valence of Ni is $d^{8-\delta}$, where $\delta$ is the hole doping.

In the large $J_H$ limit, we project to the restricted Hilbert space, which consists five states at each site\cite{PhysRevResearch.2.023112}. 
We suppress the site and layer labels for brevity here.
We label two singlon states with $\ket{\sigma}=c^{\dagger}_{1;\sigma}\ket{0}, \sigma=\uparrow,\downarrow$, 
and three doublon states with $\ket{x}=-\frac{1}{\sqrt{2}}(c^{\dagger}_{1;\uparrow}c^{\dagger}_{2;\uparrow}-c^{\dagger}_{1;\downarrow}c^{\dagger}_{2;\downarrow})\ket{0}$, 
$\ket{y}=\frac{i}{\sqrt{2}}(c^{\dagger}_{1;\uparrow}c^{\dagger}_{2;\uparrow}+c^{\dagger}_{1;\downarrow}c^{\dagger}_{2;\downarrow})\ket{0}$, and 
$\ket{z}=\frac{1}{\sqrt{2}}(c^{\dagger}_{1;\uparrow}c^{\dagger}_{2;\downarrow}+c^{\dagger}_{1;\downarrow}c^{\dagger}_{2;\uparrow})\ket{0}$. The spin operator $\vec{S}^s$ for spin-1/2 singlon and 
spin operator $\vec{S}^d$ for spin-1 doublon can be written as $(\vec{S}^s)^a=\frac{1}{2}\sum_{\sigma\sigma^{\prime}}(\sigma_a)_{\sigma\sigma^{\prime}}\ket{\sigma}\bra*{\sigma^{\prime}}$ 
and $(\vec{S}^d)^a=-i\sum_{bc}\epsilon_{abc}\ket{b}\bra*{c}$ with Pauli matrix $\sigma$ and antisymmetric tensor $\epsilon$.

We treat the kinetic terms as perturbations and perform the standard second-order perturbation theory, reaching the bilayer $t-J$ model:
\begin{eqnarray}
&&H_{t-J}=\sum_{\langle ij \rangle l\sigma}t^{\text{intra}}_{22}c^{\dagger}_{2l;i\sigma}c_{2l;j\sigma}+\sum_{il\sigma}t^{\text{inter}}_{22}c^{\dagger}_{2l;i\sigma}c_{2\bar{l};i\sigma}+H_{J}\nonumber\\
&&H_J=J_1\sum_{\langle ij \rangle l}(\vec{S}^s_{l;i}\cdot\vec{S}^s_{l;j}-\frac{1}{4}n^s_{l;i}n^s_{l;j})\nonumber\\
&&\qquad+J_2\sum_{i}(\vec{S}^s_{1;i}\cdot\vec{S}^s_{2;i}-\frac{1}{4}n^s_{1;i}n^s_{2;i})\nonumber\\
&&\qquad+J_{d1}\sum_{\langle ij\rangle l}(\vec{S}^d_{l;i}\cdot\vec{S}^d_{l;j}-n^d_{l;i}n^d_{l;j})\nonumber\\
&&\qquad+J_{d2}\sum_{i}(\vec{S}^d_{1;i}\cdot\vec{S}^d_{2;i}-n^d_{1;i}n^d_{2;i})\nonumber\\
&&\qquad+\frac{1}{2}J^{\prime}_1\sum_{\langle ij\rangle l}[(\vec{S}^s_{l;i}\cdot\vec{S}^d_{l;j}-\frac{1}{2}n^s_{l;i}n^d_{l;j})+(s \leftrightarrow d)]\nonumber\\ 
&&\qquad+\frac{1}{2}J^{\prime}_2\sum_{i}[(\vec{S}^s_{1;i}\cdot\vec{S}^d_{2;i}-\frac{1}{2}n^s_{1;i}n^d_{2;i})+(s \leftrightarrow d)]
\end{eqnarray}
where $J_1$($J_2$) is the antiferromagnetic spin coupling within(between) layers, $J_{d1}$($J_{d2}$) is the spin coupling between doublons, and $J^{\prime}_1$($J^{\prime}_2$) can be viewed as Kondo coupling between singlon and doublon.
More details of these parameters is shown in Appendix \ref{appendix A}.
In Table \ref{tab:table1}, we show the resulting spin coupling parameters with the tight-binding parameters listed above.
$n^s_{l;i}$ and $n^d_{l;i}$ are the density operator of singlon and doublon.

\begin{table}[b]
\caption{\label{tab:table1}
The spin coupling parameters for the derived effective bilayer $t-J$ model with different values of inter-layer hopping amplitude $t^{\text{inter}}_{11}$.}
\begin{ruledtabular}
\begin{tabular}{ccccccc}
  $t^{\text{inter}}_{11}$ &$J_1$ &$J_1^{\prime}$ &$J_{d1}$
  &$J_2$ &$J_2^{\prime}$ &$J_{d2}$\\
\hline
0.5& 0.046 & 0.029 & 0.072 &0.25
& 0.125 &0.05\\
0.6& 0.046 & 0.029 & 0.072 &0.36
& 0.18 &0.072 \\
0.7& 0.046 & 0.029 & 0.072 &0.49
& 0.245 &0.098\\
\end{tabular}
\end{ruledtabular}
\end{table}

We study the above $t-J$ Hamiltonian using the Gutzwiller 
approximation\cite{zhang1988renormalised,li2009gutzwiller}, 
and the trial wavefunction has the form:
\begin{eqnarray}
\ket{\psi}=\frac{\hat{P}\ket{\psi_0}}{\bra*{\psi_0}\hat{P}^2\ket{\psi_0}}
\end{eqnarray}\noindent
where $\hat{P}$ is the Gutzwiller projection operator and is assumed as:
\begin{eqnarray}
&&\hat{P}=\prod_{il} \hat{P}(il)\nonumber\\
&&\hat{P}(il)=\sum_{\sigma}\eta_{1\sigma}(il)\hat{Q}_{1e,\sigma}(il)+\sum_{\sigma\sigma^{\prime}}\eta_{2\sigma\sigma^{\prime}}(il)\hat{Q}_{2e,\sigma\sigma^{\prime}}(il)\nonumber\\
\end{eqnarray} 
here $\hat{Q}_{1e,\sigma}(il)$ is the projection operator to a singly occupied state of electron with spin $\sigma$ from orbital $1$ at site $i$ and layer $l$, 
and $\hat{Q}_{2e,\sigma\sigma^{\prime}}$ is the projection operator to a doublon state consisting of one spin $\sigma$ electron from orbital $1$ and one spin $\sigma^{\prime}$ electron from orbital $2$ at site $i$ and layer $l$.
$\eta_{1\sigma}$ and $\eta_{2\sigma\sigma^{\prime}}$ are fugacities which control the relative weight of states and ensure the local densities are same before and after the projection, that is $\langle n_{\alpha l;i\sigma}\rangle=\langle n_{\alpha l;i\sigma}\rangle_0$. 
These parameters is determined self-consistently in this work, and the explicit expressions are shown in Appendix~\ref{appendix B}.

In order to search for the minimum of energy $E=\bra*{\psi}H_{t-J}\ket{\psi}$, we first define the following expectation values: the average number $n_{\alpha l;i\sigma}=\bra*{\psi_0} c^{\dagger}_{\alpha l;i\sigma}c_{\alpha l;i\sigma}\ket{\psi_0}$, 
hopping amplitudes $\chi^{\text{intra}}_{(\alpha l;i\sigma)(\beta l;j\sigma)}=\bra*{\psi_0}c^{\dagger}_{\alpha l;i\sigma}c_{\beta l;j\sigma}\ket{\psi_0}$ and $\chi^{\text{inter}}_{(\alpha l;i\sigma)(\alpha \bar{l};i\sigma)}=\bra*{\psi_0}c^{\dagger}_{\alpha l;i\sigma}c_{\alpha \bar{l};i\sigma}\ket{\psi_0}$,
and pairing order parameter $\Delta^{\text{intra}}_{(\alpha l;i\sigma)(\alpha l;j\bar{\sigma})}=\bra*{\psi_0}c_{\alpha l;i\sigma}c_{\alpha l;j\bar{\sigma}}\ket{\psi_0}$ and $\Delta^{\text{inter}}_{(\alpha l;i\sigma)(\alpha \bar{l};i\bar{\sigma})}=\bra*{\psi_0}c_{\alpha l;i\sigma}c_{\alpha \bar{l};i\bar{\sigma}}\ket{\psi_0}$. 
We only consider the intra-orbital spin-singlet pairing and neglect the inter-orbital hopping term between layers. Site $i$ and site $j$ are the nearest neighbors.
By applying the Wick's theorem, the energy $E$ is a function of these expectation values, and the solution of $\ket{\psi_0}$ amounts to the ground state of the following effective mean-field Hamiltonian:
\begin{eqnarray}
&&H_{eff}=\sum_{\langle ij\rangle l\alpha\beta\sigma}g^t_{(\alpha l;i\sigma)(\beta l;j\sigma)}c^{\dagger}_{\alpha l;i\sigma}c_{\beta l;j\sigma}\nonumber\\
&&\ +\sum_{il\alpha\sigma}g^t_{il\alpha\sigma}c^{\dagger}_{\alpha l;i\sigma}c_{\alpha\bar{l};i\sigma}
+\sum_{il\alpha\sigma}(\frac{1}{2}\sigma h_{il\alpha}-\mu_{\alpha})c^{\dagger}_{\alpha l;i\sigma}c_{\alpha l;i\sigma}\nonumber\\
&&\ +\sum_{\langle ij\rangle l\alpha}g^I_{ijl\alpha}(c_{\alpha l;i\uparrow}c_{\alpha l;j\downarrow}-c_{\alpha l;i\downarrow}c_{\alpha l;j\uparrow})+h.c.\nonumber\\
&&\ +\sum_{i\alpha}g^I_{i\alpha}(c_{\alpha 1;i\uparrow}c_{\alpha 2;i\downarrow}-c_{\alpha 1;i\downarrow}c_{\alpha 2;i\uparrow})+h.c.,
\end{eqnarray}\noindent
where the expressions for the parameters are given by:
\begin{eqnarray}
&&g^t_{(\alpha l;i\sigma)(\beta l;j\sigma)}=\frac{\partial E}{\partial \chi^{\text{intra}}_{(\alpha l;i\sigma)(\beta l;j\sigma)}}\nonumber\\
&&g^t_{il\alpha\sigma}=\frac{\partial E}{\partial \chi^{\text{inter}}_{(\alpha l;i\sigma)(\alpha \bar{l};i\sigma)}}\nonumber\\
&&h_{il\alpha}=\frac{\partial E}{\partial n_{\alpha l;i\uparrow}}-\frac{\partial E}{\partial n_{\alpha l;i\downarrow}}\nonumber\\
&&g^I_{ijl\alpha}=\frac{\partial E}{\partial \Delta^{\text{intra}}_{(\alpha l;i \uparrow)(\alpha l;j\downarrow)}}\nonumber\\
&&g^I_{i\alpha}=\frac{\partial E}{\partial \Delta^{\text{inter}}_{(\alpha 1;i\uparrow)(\alpha 2;i\downarrow)}},
\end{eqnarray}\noindent
we adopt a self-consistent process to solve this mean-field Hamiltonian and obtain the solution of $\ket{\psi_0}$. We present our numerical results in the next section. 
\section{\label{sec:level3}The numerical result of the bilayer $t-J$ model}

\begin{figure*}
\includegraphics[width=\textwidth]{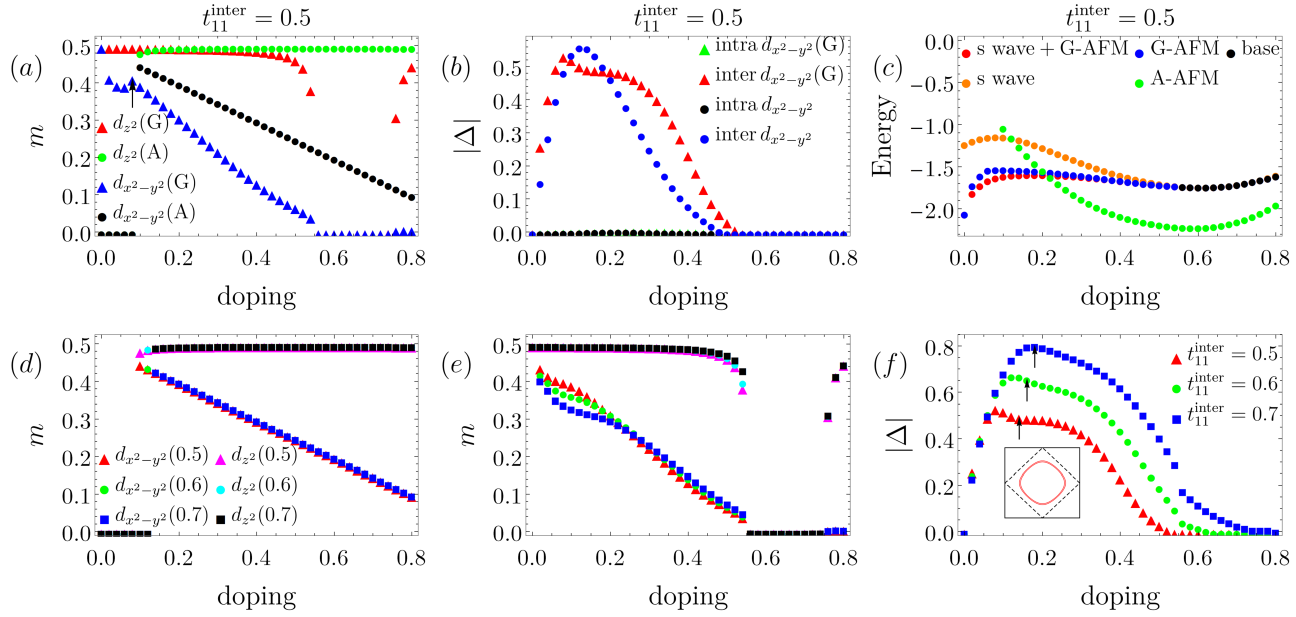}
\caption{\label{fig:FIG2}The dependence of (a) magnetic order parameter, (b) pairing order parameter and (c) energy on hole doping $\delta$ for the bilayer $t-J$ model with $t^{\text{inter}}_{11}=-0.5$.
The variations of magnetic and pairing order parameters by tuning the inter-layer hopping amplitude $t^{\text{inter}}_{11}$ are plotted in (d), (e) and (f).
(a) The magnetic order parameters of both orbitals in G-AFM phase are maximized at $\delta=0$ and gradually decrease with increasing doping. While the magnetic orders in A-AFM phase are absent in low-doping regime and appear near $\delta=0.1$.
(b) The intra-layer superconducting pairing order for both cases are negligible small, while the inter-layer pairing order in pure $s$-wave phase exhibit a dome shape and more complicated behaviors occur with an G-AFM background.
In figure (c), we demonstate the calculated energy for different states of the system. The black dots represent the states with neither magnetic order nor pairing order for comparison.
We further analyze the (d)A-AFM states and (e)(f) $s$-wave$+$G-AFM states with different $t^{\text{inter}}_{11}$, find the values of $t^{\text{inter}}_{11}$ have less impacts on the magnetic properties of the systems, but the optimal doping as well as the magnitudes of pairing order are significant changed as displayed in figure (f).}
\end{figure*}

The super-exchange couplings within layers are several times smaller than those between layers due to the large hopping parameter $t^{\text{inter}}_{11}$. 
The strong coupling of orbital $d_{z^2}$ between layers can be shared to orbital $d_{x^2-y^2}$ with large Hund's coupling, and induces finite inter-layer s-wave pairing of orbital $d_{x^2-y^2}$ in spite of small hopping for $d_{x^2-y^2}$ between layers.
This kind of pairing effectively suppress intra-layer 
cuprate-like $d$-wave pairing,
which is similar to the bilayer Hubbard models studied previously \cite{PhysRevB.80.064517,PhysRevB.84.180513} and can be verified by our numerical calculations. Meanwhile, second-order perturbation theory shows that the inter-orbital superexchange process between the half-filled and empty orbital can lead to ferromagnetic coupling\cite{PhysRevB.108.L180510,PhysRevLett.127.077204}.
So we analyze both A-AFM and G-AFM tendencies in the system and explore the possibility of the coexistence of superconductivity and magnetism.

The obtained magnetic order, superconducting order and energy as a function of hole doping are displayed in Fig.~\ref{fig:FIG2}. In the low doping limit, the intra-layer ferromagnetic order is absent until hole doping $\delta\approx 0.10$,
and this is consistent with the super-exchange picture we mentioned above, as small hole doping value leads to fewer empty orbitals.
With increasing hole doping, the system tends to form the A-AFM state, the calculated magnetic order of $d_{x^2-y^2}$ decreases linearly with hole doping and the magnetic order of $d_{z^2}$ remains at $0.5$. 
While, the G-AFM state has the largest magnetic order value at $\delta=0$, where both orbitals are half-filled and the AFM coupling determines the ground state of the system.
The strong Hund's coupling alignes the spin of the two orbitals, and causes the G-AFM orders of the two orbitals decrease simultaneously, though the filling of $d_{z^2}$ is fixed, and eventually vanish around $\delta=0.56$.
Upon increasing $t^{\text{inter}}_{11}$, the magnetic properties of the system do not change much as displayed in Fig.~\ref{fig:FIG2}(d)(e).

Fig.~\ref{fig:FIG2}(b) demonstrates the doping dependence of the calculated superconducting order parameters with and without magnetic order. 
Within the Gutwiller approximation we adopted, only the intra-orbital pairing of $d_{x^2-y^2}$ orbitals survives and the numerical calculations end up with $s$-wave spin-singlet pairing, 
different from 
the $d$-wave pairing in cuprate. 
The intra-layer pairing orders are much weaker compared to the inter-layer pairing order, since the large inter-layer super-exchange coupling of $d_{z^2}$ orbitals due to strong hopping along the vertical inter-layer bond can be shared to $d_{x^2-y^2}$ orbital.
With G-AFM order, the doping dependence of the superconducting pairing order shows more complicated feature. We analyze the Fermi surface and find the Fermi surface is fully gapped due to the presence of AFM order when doping is low, then the gap gradually decreases with increasing doping and goes to zero around $\delta=0.14$ 
[The arrows in Fig.~\ref{fig:FIG2}(f) point out the position where the Fermi surface appears, and we also gives one specific Fermi surface in the reduced magnetic Brillouin zone with $t^{\text{inter}}_{11}=-0.5$ and $\delta=0.14$].
Thus we attribute the different behaviors of superconducting pairing order parameters in different doping ranges to the change in the Fermi surfaces of this system.
We also notice that the magneic order on $d_{x^2-y^2}$ orbitals in G-AFM phase abruptly increases near $\delta=0.08$ in Fig.~\ref{fig:FIG2}(a), we believe this phenomenon is also 
caused by 
the change in Fermi surfaces.
With increasing $t^{\text{inter}}_{11}$, the value of optimal doping and magnitude of superconducting pairing order increase, which may explain the observed superconductivity in $La_{3}Ni_{2}O_{7}$ under pressure and suggest 
that electron doping this system may further enhance the superconductivity.

As displayed in Fig.~\ref{fig:FIG2}(c), the coexistence of $s$-wave superconductivity and G-AFM state has the lowest energy up to $\delta\approx 0.2$, but with a close energy for pure G-AFM state. 
Upon further increase of the hole doping, the system transfers to the A-AFM state.

As we mentioned above, the Fermi surface is fully gapped in low doping regimes, and the 
calculated pairing order parameters is too small 
to determine the onset of superconductivity.
We therefore consider the superfluid stiffness, denoted as $D_s$, to distinguish different phases of the system.
The superfluid stiffness reflects the diamagnetic Meissner effect of superconductors and is proportional to the superfluid density.
Within Kubo formulism, the superfluid stiffness is determinded by\cite{PhysRevLett.68.2830,PhysRevB.47.7995,PhysRevX.9.031049}:
\begin{eqnarray}
\frac{D_s}{\pi e^2}=\langle -k_x\rangle-\Lambda_{xx}(q_x=0,q_y\rightarrow 0,\omega=0)
\end{eqnarray}\noindent
where $k_x$ is the kinetic energy along the x direction and represents the diamagnetic responds to an external vector potential $A_x$. 
While $\Lambda_{xx}$ is the paramagnetic current-current correlation fucntion, which can be obtained from:
\begin{eqnarray}
\Lambda_{xx}(\mathbf{q},i\omega_n)=\frac{1}{N}\int_{0}^{\beta} \dd \tau e^{i\omega_n\tau}\langle j^{p}_x(\mathbf{q},\tau)j^{p}_x(-\mathbf{q},0)\rangle
\end{eqnarray}\noindent
where $\omega_n=2 \pi nT$ ($n$ is a positive integer) and the paramagnetic current $j^{p}_x(\mathbf{q})$ has the form:
\begin{eqnarray}
j^{p}_x(\mathbf{q})=\sum_{l\sigma\mathbf{k},\alpha\neq\beta} \frac{\partial t^{\text{intra}}_{22}(\mathbf{k}+\frac{\mathbf{q}}{2})}{\partial k_x} c^{\dagger}_{2l;\alpha\sigma}(\mathbf{k}+\mathbf{q})c_{2l;\beta\sigma}(\mathbf{k}) 
\end{eqnarray}\noindent
with $t^{\text{intra}}_{22}(\mathbf{k})=2t^{\text{intra}}_{22}(\cos(k_x/\sqrt{2})+\cos(k_y/\sqrt{2}))$, $\alpha(\beta)=A/B$ labels the sublattice since we are also interested in antiferromagnetic order in our work. Then we take the analytic continuation $i\omega_n\rightarrow 0+i\eta$, and use a function $A+B q_y+C q_y^2$ to fit data for small $q_y$ in order to have an appropriate $q_y\rightarrow 0$ extrapolation.
For simplicity, we approximately express the correlation function as $\Lambda_{xx}=g^2\Lambda_{xx}^0$, and the Gutzwiller renormalization factor $g$ is defined as the ratio of the expectations of nearest-neighbor intra-orbital hopping of orbital $d_{x^2-y^2}$ after and before projection, that is $g=\langle c^{\dagger}_{2l;i\sigma}c_{2l;j\sigma}\rangle/\langle c^{\dagger}_{2l;i\sigma}c_{2l;j\sigma}\rangle_0$.
Meanwhile,
\begin{eqnarray}
\Lambda_{xx}^0(\mathbf{q},i\omega_n)=\frac{1}{N}\int_{0}^{\beta} \dd \tau e^{i\omega_n\tau}\langle j^{p}_x(\mathbf{q},\tau)j^{p}_x(-\mathbf{q},0)\rangle_0 .
\end{eqnarray}\noindent

\begin{figure}[h]
\includegraphics[width=\columnwidth]{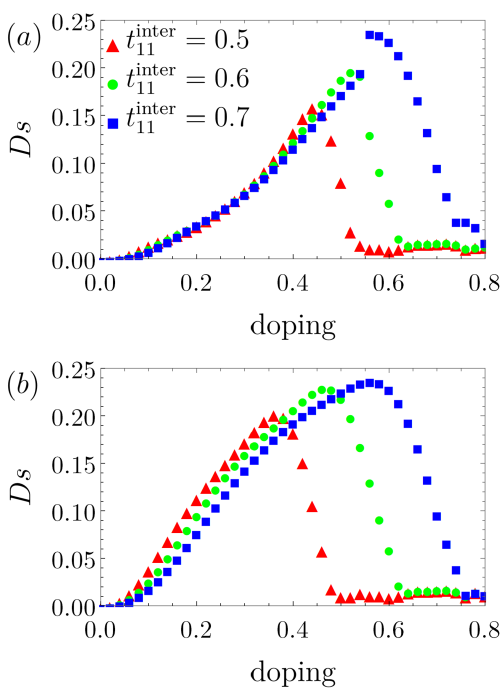}
\caption{\label{fig:FIG3}The calculated superfluid stiffness as a function of hole doping with various $t^{\text{inter}}_{11}$ in different phases. 
Superfluid stiffness (a) with and (b) without antiferromagnetic background.}
\end{figure}

In Fig.~\ref{fig:FIG3}(a), we plot the superfluid stiffness as a function of hole doping when the $s$-wave superconducting order and G-AFM order coexist.
The superfluid stiffness at low doping is largely suppressed due to the reduction of the kinetic energy after projection as well as the Gutzwiller renormalization factor.
Meanwhile, the superfluid stiffness is relatively small compared to the case without AFM background [Fig.~\ref{fig:FIG3}(b)], 
since there is no Fermi surface around $\delta=0$.
Increasing hole doping, the superfluid stiffness keeps growing as the Fermi surface gradually forms. 
After reaching the maximum value, it begins to decrease and becomes basically zero, which is consist with the result for metallic phase.
In the superconducting region, the superfluid stiffness has larger optimal doping value compared to the magnitude of pairing order, 
and the value grows from $0.44$ to $0.56$ with increasing $t^{\text{inter}}_{11}$
amplitude from $0.5$ to $0.7$.

\section{\label{sec:level4}summary}
In this work, we have performed a detailed analysis of the effective bilayer $t-J$ model using the multiband Gutzwiller approximation.
Previous theoretical studies on this material using the Gutzwiller approximation mainly focused on the smaller values of $J_H$\cite{PhysRevB.108.L140505},
and suggested large on-site intra-orbital pairing due to the pair-hopping interaction, while the intra-orbital pairing for $d_{z^2}$ orbitals on the vertical bond is subleading.
Here, we believe the nickelate $La_3Ni_2O_7$ is in the strong coupling regime and study the system in large $J_H$ limit\cite{PhysRevLett.131.206501,cao2023flat,kumar2023softening}.
Meanwhile, we carefully incorporated more inter-site correlations and derived the relatively complicated renormalized mean-field Hamiltonian, which is considered a better choice to give satisfactory results for strong correlated materials\cite{li2009gutzwiller,PhysRevB.49.12058}.   
Our numerical results show that this model prefers to form $s$-wave superconducting pairing from $d_{x^2-y^2}$ orbitals due to strong inter-layer spin exchange interaction. The superconducting dome can extend over a wide doping range, and the pairing order is enhanced with increasing 
inter-layer hopping $t^{\text{inter}}_{11}$ of $d_{z^2}$ orbitals
[see Fig.~\ref{fig:FIG3}(f)].
We note that a tensor-network analysis for a bilayer $t-J-J_{\perp}$ model shows similar results\cite{PhysRevLett.132.036502}. 
In the low hole doping regime, we find that G-AFM state can coexist with superconductivity, similar to the result for a bilayer extended Hubbard-Heisenberg model\cite{PhysRevB.106.054510}, and the 
absence of Fermi surface
causes the low value of superfluid stiffness, which will greatly suppress the superconductivity.
From DFT calculations\cite{PhysRevB.108.L180510}, the system tends to form magnetic ground state with strong electronic correlations. 
In our work, as hole doping increases, the ferromagnetic coupling within layers due to super-exchange process between half-filled and empty orbitals plays an important role, so that the A-AFM state has the lowest energy for a larger doping value.

\textit{Note added}: during the preparation of this paper, we become aware of several works which study similar models \cite{PhysRevB.108.174511,yang2023strong}.
However, we studied the interplay between magnetism and superconductivity which were not considered in those works.

\begin{acknowledgments}
This work is supported by the National Natural Science Foundation of China (No. 12274004), and the National Natural Science Foundation of China (No. 11888101).
\end{acknowledgments}

\appendix

\section{\label{appendix A}Details on the t-J model}
We constrain the Hilbert space to consist of five states at each site and perform the standard second-order perturbation theory, the superexchange coupling parameters are related to the original parameters through the following equations:
\begin{eqnarray}
&&J_1=\frac{4(t^{\text{intra}}_{11})^2}{U_1}+\frac{2(t^{\text{intra}}_{12})^2}{J_H+U^{\prime}+\epsilon_2}\nonumber\\
&&J_2=\frac{4(t^{\text{inter}}_{11})^2}{U_1}\nonumber\\
&&J_1^{\prime}=2(t^{\text{intra}}_{11})^2(\frac{1}{U_1+U^{\prime}}+\frac{1}{U_1-U^{\prime}+J_H})\nonumber\\
&&\quad+2(t^{\text{intra}}_{12})^2(\frac{1}{U_2+U^{\prime}+\epsilon_2}+\frac{1}{U_1-U^{\prime}-\epsilon_2+J_H})\nonumber\\
&&\quad+(t^{\text{intra}}_{12})^2(\frac{1}{2J_H+\epsilon_2}-\frac{1}{\epsilon_2})+\frac{(t^{intra}_{22})^2}{2J_H}\nonumber\\
&&J_2^{\prime}=2(t^{\text{inter}}_{11})^2(\frac{1}{U_1+U^{\prime}}+\frac{1}{U_1-U^{\prime}+J_H})+\frac{(t^{\text{inter}}_{22})^2}{2J_H}\nonumber\\
&&J_{d1}=\frac{(t^{\text{intra}}_{11})^2}{U_1+J_H}+\frac{(t^{\text{intra}}_{22})^2}{U_2+J_H}\nonumber\\
&&\quad+(t^{\text{intra}}_{12})^2(\frac{1}{U_1+J_H-\epsilon_2}+\frac{1}{U_2+J_H+\epsilon_2})\nonumber\\
&&J_{d2}=\frac{(t^{\text{inter}}_{11})^2}{U_1+J_H}+\frac{(t^{\text{inter}}_{22})^2}{U_2+J_H},
\end{eqnarray}\noindent
where the third term $(t^{\text{intra}}_{12})^2(\frac{1}{2J_H+\epsilon_2}-\frac{1}{\epsilon_2})$ in $J_1^{\prime}$ is from the superexchange process between the half-filled and empty orbitals,
and contributes to the intra-layer ferromagnetic coupling, which is absent in $J_2^{\prime}$ since the inter-orbital hopping between layers is zero.

\section{\label{appendix B}Derivation of the self-consistent mean-field Hamiltonian}
In the main text, we define the Gutzwiller projection operator $\hat{P}$, which is expressed in terms of fugacities $\eta$ and operators $\hat{Q}$.
To be specific, operators $\hat{Q}$ are defined as:
\begin{eqnarray}
&&\hat{Q}_{1e,\sigma}(il)=\hat{n}_{1l;i\sigma}(1-\hat{n}_{1l;i\bar{\sigma}})(1-\hat{n}_{2l;i\uparrow})(1-\hat{n}_{2l;i\downarrow})\nonumber\\
&&\hat{Q}_{2e,\sigma\sigma^{\prime}}(il)=\hat{n}_{1l;i\sigma}(1-\hat{n}_{1l;i\bar{\sigma}})\hat{n}_{2l;i\sigma^{\prime}}(1-\hat{n}_{2l;i\bar{\sigma^{\prime}}})
\end{eqnarray}\noindent
and the fugacities $\eta$ are related the expections of operators $\hat{Q}$ through:
\begin{eqnarray}
\langle \hat{Q}_{1e,\sigma}(il)\rangle&&=\frac{\langle \hat{P} \hat{Q}_{1e,\sigma}(il)\hat{P}\rangle_0 }{\langle\hat{P}\hat{P}\rangle_0}\nonumber\\
&&=\frac{\langle \prod_{jl^{\prime}\neq il}\hat{P}^2(jl^{\prime}) \hat{P}(il)\hat{Q}_{1e,\sigma}(il)\hat{P}(il)\rangle_0}{\prod_{il}z_{il}}\nonumber\\
&&=\frac{\eta_{1\sigma}^2(il)\langle \hat{Q}_{1e,\sigma}(il)\rangle_0}{z_{il}},
\end{eqnarray}
where $z_{il}=\langle\hat{P}^2(il)\rangle_0$. It is also easy to see that $\eta_{2\sigma\sigma^{\prime}}^2(il)/z_{il}=\langle\hat{Q}_{2e,\sigma\sigma^{\prime}}(il)\rangle/\langle\hat{Q}_{2e,\sigma\sigma^{\prime}}(il)\rangle_0$. 
we further make approximations about the expectation of operators $\hat{Q}$ before and after the projection, so that:
\begin{eqnarray}
&&\langle\hat{Q}_{1e,\sigma}(il)\rangle=\delta \,n_{1l;i\sigma}\nonumber\\
&&\langle\hat{Q}_{2e,\sigma\sigma^{\prime}}(il)\rangle=n_{1l;i\sigma}n_{2l;i\sigma^{\prime}}\nonumber\\
&&\langle\hat{Q}_{1e,\sigma}(il)\rangle_0=n_{1l;i\sigma}^0(1-n_{1l;i\bar{\sigma}}^0)(1-n_{2l;i\uparrow}^0)(1-n_{2l;i\downarrow}^0)\nonumber\\
&&\langle\hat{Q}_{2e,\sigma\sigma^{\prime}}(il)\rangle_0=n_{1l;i\sigma}^0(1-n_{1l;i\bar{\sigma}}^0)n_{2l;i\sigma^{\prime}}^0(1-n_{2l;i\bar{\sigma^{\prime}}}^0)\nonumber\\
\end{eqnarray}\noindent
Where $\delta$ is the hole doping parameter. Here, we assume the electron densities before and after projection are same, that is $n_{\alpha l;i\sigma}=n_{\alpha l;i\sigma}^0$. With these relations, we can evaluate the energy $E=\langle \hat{H}_{t-J}\rangle$. 
By taking Wick contractions, the energy $E$ can be express in terms of expectation values with respect to the non-interacting state $\ket{\psi_0}$. Here we take one kinetic term as an example to illustrate the process.
\begin{eqnarray}
&&\langle c^{\dagger}_{2l;i\sigma}c_{2l;j\sigma}\rangle\nonumber\\
&&=\frac{\langle\hat{P}(il)c^{\dagger}_{2l;i\sigma}\hat{P}(il)\hat{P}(jl)c_{2l;j\sigma}\hat{P}(jl)\rangle_0}{z_{il}z_{jl}}\nonumber\\
&&=\frac{1}{z_{il}z_{jl}}[\eta_{1\sigma}(il)\eta_{2\sigma\sigma}(il)\eta_{1\sigma}(jl)\eta_{2\sigma\sigma}(jl)\nonumber\\
&&\quad\quad \langle \hat{Q}_{2e,\sigma\sigma}(il) c^{\dagger}_{2l;i\sigma}\hat{Q}_{1e,\sigma}(il)\hat{Q}_{1e,\sigma}(jl)c_{2l;j\sigma}\hat{Q}_{2e,\sigma\sigma}(jl)\rangle_0\nonumber\\
&&\quad\quad+\dots]
\end{eqnarray}\noindent
Considering the Hamiltonian is defined in the restricted Hilbert space, the operator $c^{\dagger}_{2l;i\sigma}$ should be understanded as $c^{\dagger}_{2l;i\sigma}\equiv\hat{P}_{triplet}(il)c^{\dagger}_{2l;i\sigma}$, with $\hat{P}_{triplet}(il)=(\vec{S}_{1l;i}\cdot\vec{S}_{2l;i}+\frac{3}{4})$ projecting out the spin-singlet state.
Similarly $c_{2l;j\sigma}\equiv c_{2l;j\sigma}\hat{P}_{triplet}(jl)$.
We substitute $(B1)$ into equation $(B4)$, the first term in $(B4)$ is then:
\begin{eqnarray}
&&\langle \hat{Q}_{2e,\sigma\sigma}(il) c^{\dagger}_{2l;i\sigma}\hat{Q}_{1e,\sigma}(il)\hat{Q}_{1e,\sigma}(jl)c_{2l;j\sigma}\hat{Q}_{2e,\sigma\sigma}(jl)\rangle_0\nonumber\\
&&=-\langle c^{\dagger}_{1l;i\sigma}c^{\dagger}_{1l;j\sigma}c^{\dagger}_{2l;i\sigma}c_{1l;i\sigma}c_{1l;j\sigma}c_{2l;j\sigma}\rangle_0+\dots\nonumber\\
&&=n_{1l;i\sigma}n_{1l;j\sigma}\chi^{\text{intra}}_{(2l;i\sigma)(2l;j\sigma)}\nonumber\\
&&\quad-\chi^{\text{intra}}_{(1l;i\sigma)(1l;j\sigma)}\chi^{\text{intra}}_{(1l;j\sigma)(1l;i\sigma)}\chi^{\text{intra}}_{(2l;i\sigma)(2l;j\sigma)}\nonumber\\
&&\quad+\chi^{\text{intra}}_{(1l;i\sigma)(2l;j\sigma)}\chi^{\text{intra}}_{(1l;j\sigma)(1l;i\sigma)}\chi^{\text{intra}}_{(2l;i\sigma)(1l;j\sigma)}+\dots
\end{eqnarray}
After evaluating the J-term as was done for the kinetic term above, we finally get the total energy $E=E(\chi,\Delta,n)$ which is expressed in terms of non-interacting expectations.
In this work, we don't relate the expectations after projection to the expectations before projection by a simple multiplicative factor, like $\langle c^{\dagger}_{2l;i\sigma}c_{2l;j\sigma}\rangle=g_t \langle c^{\dagger}_{2l;i\sigma}c_{2l;j\sigma}\rangle_0$, 
instead, we include the inter-site correlation and take all possible Wick contractions to derive a relatively complicated expression for the energy. 
We believe that this is a better choice to deal with the strong correlated materials.

\section{\label{appendix C}Calculation of the superfluid stiffness}
With the self-consistent numerical results of the mean-field Hamiltonian, it is straighforward to evaluate the equation for superfluid stiffness  presented in the main text.
Substituting equation $(9)$ into $(8)$, the resulting current-current correlation function can be expressed as:
\begin{figure*}
\includegraphics[width=\textwidth]{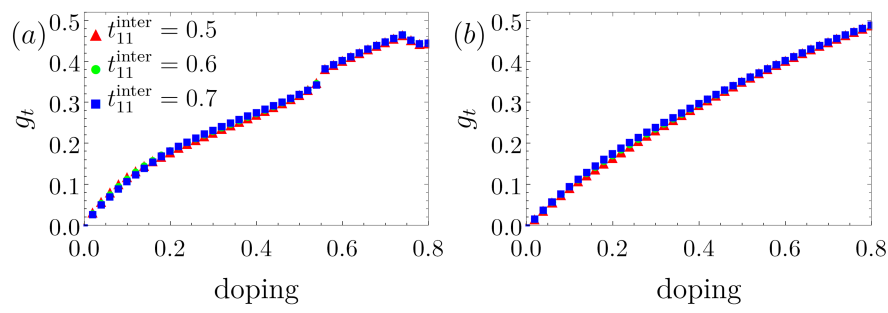}
\caption{\label{fig:Fig4} The renormalized factor $g$ versus hole doping for superconducting phases (a) with and (b) without G-AFM order.}
\end{figure*}
\begin{widetext}
  \begin{eqnarray}
    &&\quad \Lambda_{xx}(\mathbf{q},i\omega_n)\nonumber\\
    &&=C\int_0^{\beta} d\tau e^{i\omega_n\tau}\sum_{\substack{ll^{\prime}\sigma\sigma^{\prime}\mathbf{k}\mathbf{k}^{\prime} \\ \alpha\neq\beta ,\gamma\neq\delta}}\sin(\frac{k_x+\frac{q_x}{2}}{\sqrt{2}})\sin(\frac{k_x^{\prime}-\frac{q_x}{2}}{\sqrt{2}})
    \langle c^{\dagger}_{2l;\alpha\sigma}(\mathbf{k}+\mathbf{q},\tau)c_{2l;\beta\sigma}(\mathbf{k},\tau)c^{\dagger}_{2l^{\prime};\gamma\sigma^{\prime}}(\mathbf{k}^{\prime}-\mathbf{q},0)c_{2l^{\prime},\delta\sigma^{\prime}}(\mathbf{k}^{\prime},0)\rangle\nonumber\\
    &&=Cg^2\int_0^{\beta} d\tau e^{i\omega_n\tau}\sum_{\substack{ll^{\prime}\sigma\sigma^{\prime}\mathbf{k}\mathbf{k}^{\prime} \\ \alpha\neq\beta, \gamma\neq\delta}}\sin(\frac{k_x+\frac{q_x}{2}}{\sqrt{2}})\sin(\frac{k_x^{\prime}-\frac{q_x}{2}}{\sqrt{2}})(A+B)
  \end{eqnarray}
  where $C=2(t^{\text{intra}}_{22})^2/N$, and the elements $A$ and $B$ are given by:
  \begin{eqnarray}
    &&A=-\langle c^{\dagger}_{2l;\alpha\sigma}(\mathbf{k}+\mathbf{q},\tau)c^{\dagger}_{2l^{\prime};\gamma\sigma^{\prime}}(\mathbf{k}^{\prime}-\mathbf{q},0)\rangle_0 \langle c_{2l;\beta\sigma}(\mathbf{k},\tau)c_{2l^{\prime},\delta\sigma^{\prime}}(\mathbf{k}^{\prime},0)\rangle_0\\
    &&B=\langle c^{\dagger}_{2l;\alpha\sigma}(\mathbf{k}+\mathbf{q},\tau)c_{2l^{\prime},\delta\sigma^{\prime}}(\mathbf{k}^{\prime},0)\rangle_0\langle c_{2l;\beta\sigma}(\mathbf{k},\tau)c^{\dagger}_{2l^{\prime};\gamma\sigma^{\prime}}(\mathbf{k}^{\prime}-\mathbf{q},0)\rangle_0
  \end{eqnarray}
\end{widetext}

In the last step of equation $(C1)$, we apply Wick's theorem. Using the unitary matrix $U(\mathbf{k})$ which diagonalizes the effective mean-field Hamiltonian in momentum space with the basis operator 
$\Psi ^{\dagger}_{Sl;\alpha\sigma}(\mathbf{k})=[c^{\dagger}_{1t;A\uparrow}(\mathbf{k}),c^{\dagger}_{1t;B\uparrow}(\mathbf{k}),c^{\dagger}_{2t;A\uparrow}(\mathbf{k}),c^{\dagger}_{2t;B\uparrow}(\mathbf{k}),c^{\dagger}_{1b;A\uparrow}(\mathbf{k}),c^{\dagger}_{1b;B\uparrow}(\mathbf{k}),\\c^{\dagger}_{2b;A\uparrow}(\mathbf{k}),c^{\dagger}_{2b;B\uparrow}(\mathbf{k}),
c_{1t;A\downarrow}(-\mathbf{k}),c_{1t;B\downarrow}(-\mathbf{k}),c_{2t;A\downarrow}(-\mathbf{k}),\\c_{2t;B\downarrow}(-\mathbf{k}),c_{1b;A\downarrow}(-\mathbf{k}),c_{1b;B\downarrow}(-\mathbf{k}),c_{2b;A\downarrow}(-\mathbf{k}),c_{2b;B\downarrow}(-\mathbf{k})]$, the correlators in the above equation can be further expressed as:
\begin{widetext}
  \begin{eqnarray}
    &&\quad \sum_{\sigma\sigma^{\prime}}\langle c^{\dagger}_{2l;\alpha\sigma}(\mathbf{k}+\mathbf{q},\tau)c^{\dagger}_{2l^{\prime};\gamma\sigma^{\prime}}(\mathbf{k}^{\prime}-\mathbf{q},0)\rangle_0 \langle c_{2l;\beta\sigma}(\mathbf{k},\tau)c_{2l^{\prime},\delta\sigma^{\prime}}(\mathbf{k}^{\prime},0)\rangle_0\nonumber\\
    &&=-\delta_{\mathbf{k}^{\prime},-\mathbf{k}}\sum_{n,m}U^{\dagger n}_{(2l;\alpha\uparrow)}(\mathbf{k}+\mathbf{q})U^{(2l^{\prime};\gamma\downarrow)}_{n}(\mathbf{k}+\mathbf{q}) U^{\dagger m}_{(2l^{\prime};\delta\downarrow)}(\mathbf{k})
    U^{(2l;\beta\uparrow)}_{m}(\mathbf{k})\langle \tilde{c}^{\dagger}_{n\mathbf{k}+\mathbf{q}}(\tau)\tilde{c}_{n\mathbf{k}+\mathbf{q}}(0)\rangle_0\langle \tilde{c}^{\dagger}_{m\mathbf{k}}(0)\tilde{c}_{m\mathbf{k}}(\tau)\rangle_0\nonumber\\
    &&\quad+U^{\dagger n}_{(2l^{\prime};\gamma\uparrow)}(\mathbf{k}^{\prime}-\mathbf{q})U^{(2l;\alpha\downarrow)}_{n}(\mathbf{k}^{\prime}-\mathbf{q}) U^{\dagger m}_{(2l;\beta\downarrow)}(\mathbf{k}^{\prime})
    U^{(2l^{\prime};\delta\uparrow)}_{m}(\mathbf{k}^{\prime})\langle \tilde{c}^{\dagger}_{n\mathbf{k}^{\prime}-\mathbf{q}}(0)\tilde{c}_{n\mathbf{k}^{\prime}-\mathbf{q}}(\tau)\rangle_0\langle \tilde{c}^{\dagger}_{m\mathbf{k}^{\prime}}(\tau)\tilde{c}_{m\mathbf{k}^{\prime}}(0)\rangle_0\nonumber\\[10pt]
    &&\quad\sum_{\sigma\sigma^{\prime}}\langle c^{\dagger}_{2l;\alpha\sigma}(\mathbf{k}+\mathbf{q},\tau)c_{2l^{\prime};\delta\sigma^{\prime}}(\mathbf{k}^{\prime},0)\rangle_0 \langle c_{2l;\beta\sigma}(\mathbf{k},\tau)c^{\dagger}_{2l^{\prime},\gamma\sigma^{\prime}}(\mathbf{k}^{\prime}-\mathbf{q},0)\rangle_0\nonumber\\
    &&=-\delta_{\mathbf{k}^{\prime},\mathbf{k}+\mathbf{q}}\sum_{n,m}U^{\dagger n}_{(2l;\alpha\uparrow)}(\mathbf{k}+\mathbf{q})U^{(2l^{\prime};\delta\uparrow)}_{n}(\mathbf{k}+\mathbf{q}) U^{\dagger m}_{(2l^{\prime};\gamma\uparrow)}(\mathbf{k})
    U^{(2l;\beta\uparrow)}_{m}(\mathbf{k})\langle \tilde{c}^{\dagger}_{n\mathbf{k}+\mathbf{q}}(\tau)\tilde{c}_{n\mathbf{k}+\mathbf{q}}(0)\rangle_0\langle \tilde{c}^{\dagger}_{m\mathbf{k}}(0)\tilde{c}_{m\mathbf{k}}(\tau)\rangle_0\nonumber\\
    &&\quad+U^{\dagger n}_{(2l^{\prime};\delta\downarrow)}(-\mathbf{k}^{\prime})U^{(2l;\alpha\downarrow)}_{n}(-\mathbf{k}^{\prime}) U^{\dagger m}_{(2l;\beta\downarrow)}(-\mathbf{k}^{\prime}+\mathbf{q})
    U^{(2l^{\prime};\gamma\downarrow)}_{m}(-\mathbf{k}^{\prime}+\mathbf{q})\langle \tilde{c}^{\dagger}_{n-\mathbf{k}^{\prime}}(0)\tilde{c}_{n-\mathbf{k}^{\prime}}(\tau)\rangle_0\langle \tilde{c}^{\dagger}_{m-\mathbf{k}^{\prime}+\mathbf{q}}(\tau)\tilde{c}_{m-\mathbf{k}^{\prime}+\mathbf{q}}(0)\rangle_0\nonumber\\
  \end{eqnarray}
  where $\tilde{c}^{\dagger}_{n\mathbf{k}}$ are defined as $\tilde{c}^{\dagger}_{n\mathbf{k}}=c^{\dagger}_{Sl;\alpha\sigma}(\mathbf{k})U^{(Sl;\alpha\sigma)}_n(\mathbf{k})$. Considering:
  \begin{eqnarray}
    &&\langle \tilde{c}^{\dagger}_{n\mathbf{k}}(\tau)\tilde{c}_{n\mathbf{k}}(0)\rangle_0=e^{E_n(\mathbf{k})\tau}n_F(E_n(\mathbf{k}))\nonumber\\
    &&\langle \tilde{c}^{\dagger}_{n\mathbf{k}}(0)\tilde{c}_{n\mathbf{k}}(\tau)\rangle_0=-(1-n_F(E_n(\mathbf{k})))e^{-E_n(\mathbf{k})\tau}
  \end{eqnarray}
  With these relations, we finally have:
  \begin{eqnarray}
    &&\quad \Lambda_{xx}(\mathbf{q},i\omega_n)\nonumber\\
    &&=Cg^2\sum_{\substack{ll^{\prime}\mathbf{k}nm\\ \alpha\neq\beta,\gamma\neq\delta}}[-\sin(\frac{k_x+\frac{q_x}{2}}{\sqrt{2}})^2 U^{\dagger n}_{(2l;\alpha\uparrow)}(\mathbf{k}+\mathbf{q})U^{(2l^{\prime};\gamma\downarrow)}_{n}(\mathbf{k}+\mathbf{q})U^{\dagger m}_{2l^{\prime};\delta\downarrow}(\mathbf{k})U^{(2l;\beta\uparrow)}_m(\mathbf{k})\frac{n_F(E_n(\mathbf{k}+\mathbf{q}))-n_F(E_m(\mathbf{k}))}{i\omega_n+E_n(\mathbf{k}+\mathbf{q})-E_m(\mathbf{k})}\nonumber\\
    &&\quad \quad \quad  -\sin(\frac{k_x-\frac{q_x}{2}}{\sqrt{2}})^2 U^{\dagger n}_{(2l^{\prime};\gamma\uparrow)}(\mathbf{k}-\mathbf{q})U^{(2l;\alpha\downarrow)}_{n}(\mathbf{k}-\mathbf{q})U^{\dagger m}_{2l;\beta\downarrow}(\mathbf{k})U^{(2l^{\prime};\delta\uparrow)}_m(\mathbf{k})\frac{n_F(E_m(\mathbf{k}))-n_F(E_n(\mathbf{k}-\mathbf{q}))}{i\omega_n+E_m(\mathbf{k})-E_n(\mathbf{k}-\mathbf{q})}\nonumber\\
    &&\quad \quad \quad  -\sin(\frac{k_x+\frac{q_x}{2}}{\sqrt{2}})^2 U^{\dagger n}_{(2l;\alpha\uparrow)}(\mathbf{k}+\mathbf{q})U^{(2l^{\prime};\delta\uparrow)}_{n}(\mathbf{k}+\mathbf{q})U^{\dagger m}_{2l^{\prime};\gamma\uparrow}(\mathbf{k})U^{(2l;\beta\uparrow)}_m(\mathbf{k})\frac{n_F(E_n(\mathbf{k}+\mathbf{q}))-n_F(E_m(\mathbf{k}))}{i\omega_n+E_n(\mathbf{k}+\mathbf{q})-E_m(\mathbf{k})}\nonumber\\
    &&\quad \quad \quad  -\sin(\frac{k_x-\frac{q_x}{2}}{\sqrt{2}})^2 U^{\dagger n}_{(2l^{\prime};\delta\downarrow)}(-\mathbf{k})U^{(2l;\alpha\downarrow)}_{n}(-\mathbf{k})U^{\dagger m}_{2l;\beta\downarrow}(-\mathbf{k}+\mathbf{q})U^{(2l^{\prime};\gamma\downarrow)}_m(-\mathbf{k}+\mathbf{q})\frac{n_F(E_m(-\mathbf{k}+\mathbf{q}))-n_F(E_n(-\mathbf{k}))}{i\omega_n+E_m(-\mathbf{k}+\mathbf{q})-E_n(-\mathbf{k})}]\nonumber\\
  \end{eqnarray}
\end{widetext}
The kinetic energy term for the superfluid stiffness can be obtained according to the method in Appendix~\ref{appendix B},
and we use it to calculate the renormalized factor $g$. The corresponding factors for different states are shown Fig.~\ref{fig:Fig4}.
\nocite{*}

\bibliography{nickel}

\end{document}